\documentclass[twocolumn,prb,floatfix,citeautoscript,cite]{revtex4}

\usepackage{graphicx}
\usepackage{epstopdf}
\usepackage{color}
\usepackage{amsmath}

\begin{document}

\title{Hematite at its thinnest limit}

\author{C. Bacaksiz}
\affiliation{Theory of Functional Materials, Department of Physics, University of Antwerp, Groenenborgerlaan
171, B-2020 Antwerp, Belgium}
\affiliation{NANOlab Center of Excellence, University of Antwerp, Belgium}

\author{M. Yagmurcukardes}
\affiliation{Theory of Functional Materials, Department of Physics, University of Antwerp, Groenenborgerlaan
171, B-2020 Antwerp, Belgium}
\affiliation{NANOlab Center of Excellence, University of Antwerp, Belgium}

\author{F. M. Peeters}
\affiliation{Theory of Functional Materials, Department of Physics, University of Antwerp, Groenenborgerlaan
171, B-2020 Antwerp, Belgium}
\affiliation{NANOlab Center of Excellence, University of Antwerp, Belgium}

\author{M. V. Milo\v{s}evi\'c}
\email{milorad.milosevic@uantwerpen.be}
\affiliation{Theory of Functional Materials, Department of Physics, University of Antwerp, Groenenborgerlaan
171, B-2020 Antwerp, Belgium}
\affiliation{NANOlab Center of Excellence, University of Antwerp, Belgium}

\begin{abstract}
Motivated by the recent synthesis of two-dimensional $\alpha$-Fe$_2$O$_3$ 
[Balan \textit{et al.} Nat. Nanotech. \textbf{13}, 602 (2018)], we analyze the 
structural, vibrational, electronic and magnetic properties of single- and 
few-layer $\alpha$-Fe$_2$O$_3$ compared to bulk, by \textit{ab-initio} and 
Monte-Carlo simulations. We reveal how monolayer $\alpha$-Fe$_2$O$_3$ (hematene) 
can be distinguished from the few-layer structures, and how they all differ from 
bulk through observable Raman spectra. The optical spectra exhibit gradual shift 
of the prominent peak to higher energy, as well as additional features at lower 
energy when $\alpha$-Fe$_2$O$_3$ is thinned down to a monolayer. Both optical 
and electronic properties have strong spin asymmetry, meaning that lower-energy 
optical and electronic activities are allowed for the single-spin state. 
Finally, our considerations of magnetic properties reveal that 2D hematite has 
anti-ferromagnetic ground state for all thicknesses, but the critical 
temperature for Morin transition increases with decreasing sample thickness. On all accounts, the link to available experimental data is made, and further measurements are prompted. 
\end{abstract}
\maketitle

\section{Introduction}

Following the successful isolation of graphene\cite{Novo1,Geim1}, an immense 
effort has been exerted on the synthesis of other two dimensional (2D) 
layered materials such as group III-V binary compounds ($h$-BN, 
$h$-AlN)\cite{Sahin3,Wang2,Kim,Tsipas,Bacaksiz} and transition-metal 
dichalcogenides 
(TMDs)\cite{Gordon,Coleman,Wang1,Ross,Sahin2,Tongay,Horzum,Chen3}. Most of these 
2D materials have bulk counterparts that are formed by van der Waals (vdW) 
stacked layers. Beside those, the atomically thin layers of 
silicene\cite{Cahangirov,Kara} and germanene\cite{Cahangirov}, which have 
non-layered bulk forms, have been successfully synthesized. Over the last years, 
the synthesis of ultra-thin materials from their non-layered bulk counterparts 
has 
also attracted interest in the material science\cite{Ga_N6,ga1,ga2}, where one 
of the most recent successes is the extraction of ultra-thin $\alpha$-Fe$_2$O$_3$ (2D hematite)\cite{Balan}.

Iron oxide is a well-studied material family including several structural 
phases. The family exhibits various electronic and magnetic properties, such as 
band 
gap ranging from insulator to semiconductor\cite{Kan,Marelli,Mishra,Chen,Zeng}, 
super-paramagnetism\cite{Katz}, and weak 
ferromagnetism\cite{Morin,Moriya,Sakurai,Namai,Ohkoshi,Namai2}. Iron oxide has 
four different crystalline phases at ambient pressures: $\alpha$-Fe$_2$O$_3$ 
(hematite), $\beta$-Fe$_2$O$_3$, $\gamma$-Fe$_2$O$_3$ (maghemite), and 
$\varepsilon$-Fe$_2$O$_3$. Among these phases, hematite is a thermodynamically 
stable polymorph of Fe$_2$O$_3$, and a well-known ferromagnet that undergoes 
magnetic phase transitions from paramagnetic to weak-ferromagnetic state at the 
Neel temperature ($T_{N}$) of 961 K, and from weak-ferromagnetic to 
anti-ferromagnetic state at Morin transition temperature ($T_{M}$) of 265 
K\cite{Morin,Moriya}. In addition, it has optimal band gap for light absorption 
applications\cite{Smith}, and has been widely investigated for several other 
technological applications such as gas sensing\cite{Chen,Long}, lithium-ion 
batteries\cite{Wu,Banerjee,Gu}, water treatment\cite{Tang,Cao,Wang}, and 
catalysis\cite{Qiu}. 

It is already well established that nano-structuring of a material can 
significantly influence its magnetic and other properties due to quantum 
confinement and surface effects\cite{Dormann}. Such effects on 
$\alpha$-Fe$_2$O$_3$ were also studied. Schroeer \textit{et al.} reported that 
$T_{M}$ of microcrystal hematite is depressed under negative pressure induced 
due to the lattice spacing\cite{Schroeer}. Zysler \textit{et al.} demonstrated 
that $T_{M}$ increases as the size of hematite nanoparticles increases\cite{Zysler}. In addition, Sorescu \textit{et al.} studied the weak 
ferromagnetic phase above $T_{M}$ and an anti-ferromagnetic phase below $T_{M}$ 
for different sizes and morphologies of hematite nanoparticles\cite{Sorescu}. 
Only recently, the focus of research shifted to the effects of thickness, after 
Balan \textit{et al.} synthesized atomically-thin hematite and reported that 
it exhibits optical band gap comparable with the bulk sample, 
and weak ferromagnetism at low temperature regime without any Morin 
transition\cite{Balan}. In their report, two different 2D hematite samples were considered, exfoliated from [001] and [010] crystallographic directions, and [001] 2D samples were claimed to be more stable based on the molecular dynamics simulations. They also attempted to support the experimental results 
by a DFT study, however, they considered only on a monolayer structure, which was not present in experiment at all. In the meantime, 2D $\alpha$-Fe$_2$O$_3$ (named hematene) was reported in experimental and in DFT-based study as a stable 
anti-ferromagnetic semiconductor,\cite{Gonzalez,Padilha} same as bulk, which 
further clouds the 
conclusions regarding thickness dependence of the magnetic properties of 
hematite in its thinnest limit.

Motivated by above issues, and recognizing the importance of hematene as a 
newest member of the family of magnetic 2D materials (rapidly emerging after the 
synthesis of the first magnetic single-layer of CrI$_{3}$\cite{Huang}) that may 
have many other uses as well, we here thoroughly investigate its 
thickness-dependent physical properties, namely structural, vibrational, 
electronic, and magnetic properties of monolayer, few-layer, and bulk 
$\alpha$-Fe$_2$O$_3$ by performing density functional theory-based calculations 
and Monte-Carlo simulations. In doing so, we reveal how the atomistic thickness 
of $\alpha$-Fe$_2$O$_3$ is clearly reflected in the Raman and visible range 
optical spectra, and the critical 
temperature, all relevant to the ongoing experimental efforts in the field.

The paper is organized as follows. Details of the computational methodology are presented in Sec. \ref{comp}. The thickness-dependent structural, vibrational, electronic, optical, and magnetic properties of hematene versus few-layer and bulk hematite are then comparatively discussed in Secs. 
\ref{struct}-\ref{magnet}. We summarize our results and conclusions in Sec. 
\ref{Conc}.

\section{Computational methodology}\label{comp}

To investigate the structural, vibrational, magnetic, and electronic, properties 
of a two-dimensional $\alpha$-Fe$_2$O$_3$ crystal, first-principle calculations 
were performed in the framework of density functional theory (DFT) as 
implemented in the Vienna \textit{ab-initio} simulation package 
(VASP)\cite{vasp1,vasp2}. The Perdew-Burke-Ernzerhof (PBE)\cite{perdew} form of the generalized gradient approximation (GGA) was adopted to describe electron exchange and correlation. The Hubbard $U$ term was included to be 4 eV for Fe atom in all calculations to account for the strong on-site Coulomb 
interaction\cite{Dudarev}. The van der Waals (vdW) correction to the GGA 
functional was included by using the DFT-D2 method of Grimme\cite{grimme}. The charge transfer in the system was determined by the 
Bader technique\cite{bader}.

The kinetic energy cut-off for plane-wave expansion was set to 600 eV and the 
energy was minimized until its variation in the following steps became lower 
than 10$^{-6}$ eV. The Gaussian smearing method was employed for the total 
energy calculations. In order to capture the correct occupancy of the surface states, the width of smearing was set to 0.01 eV (checked to be sufficiently small for few-layer Fe$_{2}$O$_{3}$ samples). Total 
Hellmann-Feynman forces was taken to be 10$^{-5}$ eV/\AA~ for the structural 
optimization. $24\times24\times1$ $\Gamma$-centered \textit{k}-point sampling 
was used in the primitive unit cells. To avoid interaction between successive 
layers in the out-of-plane direction, we used a vacuum spacing of 12 \AA. Regarding the dielectric function calculations, $48\times48\times1$ $\Gamma$-centered \textit{k}-point sampling was used and direct ($q=0$) single-particle excitations from valence to conduction band were calculated. The local field effects were included at DFT level.

In order to investigate the dynamical stability of a monolayer hematene 
structure, the phonon band dispersions were calculated by the small displacement 
method, as implemented in the PHON code\cite{phon}. In addition, the first order 
off-resonant Raman activities of the phonon modes at the $\Gamma$ point were 
obtained by calculating the change in the macroscopic dielectric tensor with 
respect to the normal mode describing each vibrational mode using 
finite-difference method\cite{vasp-raman}. A detailed theory of the Raman 
scattering can be found in our recent studies\cite{raman-myk,r1}.

\begin{figure*}[t]
\includegraphics[width=0.8\linewidth]{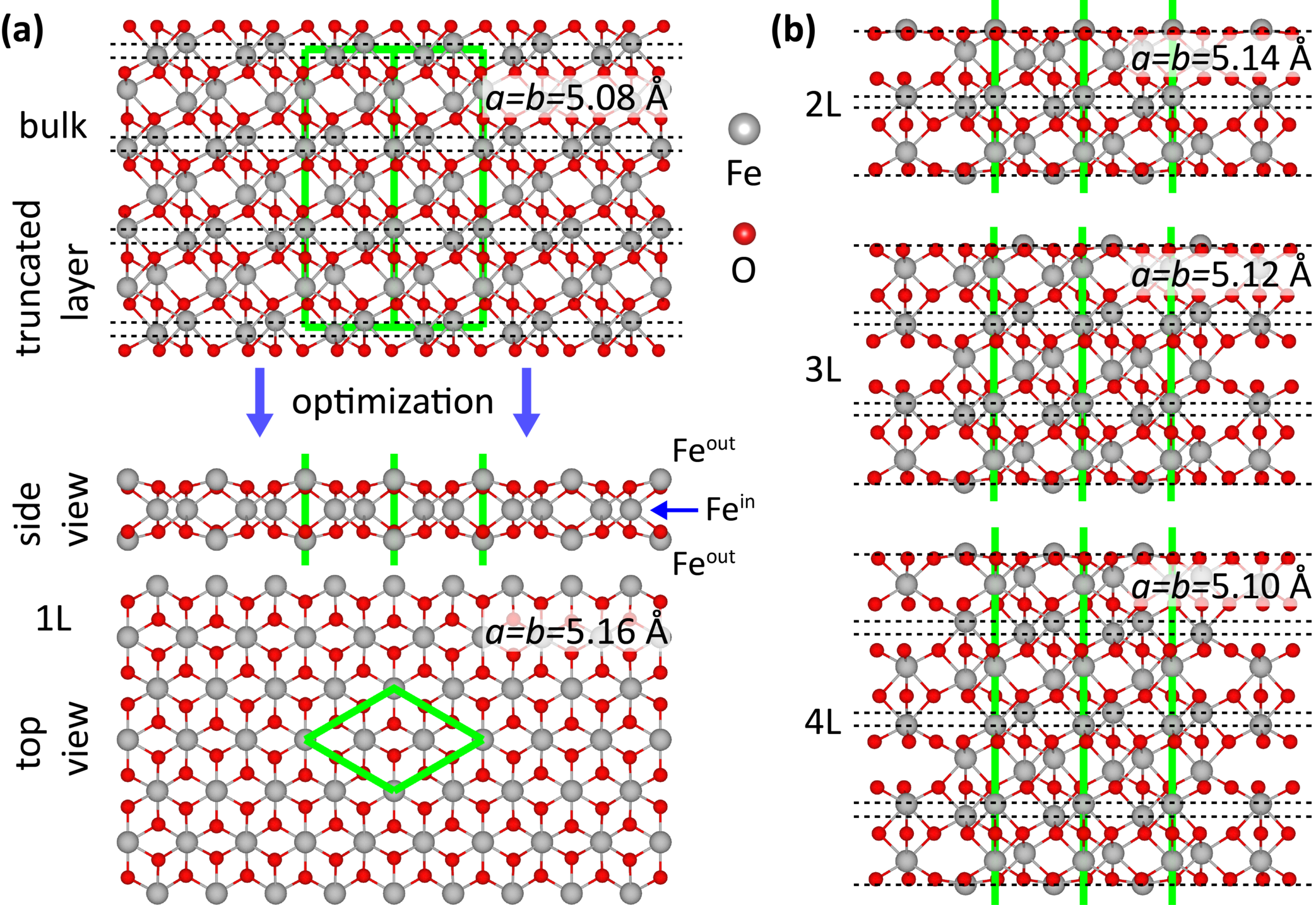}
\caption{\label{f1}
(color online) (a) The side and top views of hematene truncated from the bulk 
$\alpha$-Fe$_2$O$_3$ in the [001] direction, after optimization. Green line 
denotes the primitive unit cell. Two types of Fe atoms are also indicated as 
Fe$^{\text{in}}$ and 
Fe$^{\text{out}}$. (b) The side view of the optimized structure of 2L, 3L, and 4L $\alpha$-Fe$_2$O$_3$. } 
\end{figure*}

To investigate the magnetic exchange parameters between the magnetic sites, we used four-state methodology\cite{Xiang}, which relies on mapping of the 
energetics of different magnetic configurations onto Heisenberg 
spin Hamiltonian:
\begin{align}\label{ham}
 H = \sum_{ \langle  ij \rangle_{n}  } -J_{n}^{ij} S_{i}\cdot 
S_{j},
\end{align}
where $J_{n}^{ij}$ is the magnetic exchange parameter between $i$th and $j$th magnetic 
sites, indexed by $n$ in the nearest-neighbor sequence. Considering two atomic sites, 1 and 2, which are the nearest neighbors of each other and the corresponding exchange parameter, $J_{1}^{12}$, the energy can be written as:
\begin{align}\label{spineng}
 E = J_{1}^{12} S_{1} \cdot S_{2} + S_{1}\cdot \text{K$_1$}  + 
S_{2}\cdot \text{K$_2$}  + E_{others}.
\end{align} 
Here the first term describes the interaction between sites 1 and 2, $\text{K$_1$}= \textstyle \sum_{ \langle  1j\neq1,2 \rangle_{n}  } J_{n}^{1j} S_{1}$ and 
$\text{K$_2$}= \textstyle \sum_{ \langle  2j\neq1,2 \rangle_{n}  } J_{n}^{2j} S_{2}$ are the interactions of site 1 with others and site 2 with others, respectively. Last term stands for the interaction between the 
sites different from 1 and 2. To isolate $J_{1}^{12}$, four different magnetic configurations were chosen as (i) $S_1 = S, S_2 =S$; (ii) $S_1=-S, S_2=S$; (iii) $S_1=S, S_2=-S$; (iv) $S_1=-S, S_2=-S$, and $S_{other}=S$, so that
\begin{align}\label{states}
&E^{\text{(i)}}   = E_{others}+J_{1}^{12} S^{2} + S^{2} \text{$K_1$} + {S}^{2} \text{$K_2$}, \nonumber \\ \nonumber
&E^{\text{(ii)}}  = E_{others}-J_{1}^{12} S^{2} - S^{2} \text{$K_1$} + {S}^{2} \text{$K_2$}, \\ \nonumber
&E^{\text{(iii)}} = E_{others}-J_{1}^{12} S^{2} + S^{2} \text{$K_1$} - {S}^{2} \text{$K_2$}, \\ \nonumber
&E^{\text{(iv)}}  = E_{others}+J_{1}^{12} S^{2} - S^{2} \text{$K_1$} - {S}^{2} \text{$K_2$}.
\end{align} 
This set of equations yields $J_{1}^{12}$ as
\begin{align}
 J_{1}^{12}=\frac{E^{\text{(i)}}+E^{\text{(iv)}}-E^{\text{(ii)}}-E^{\text{(iii)}}}{4S^{2}},
\end{align}
which is a general formula for any considered pair of magnetic sites. 

The temperature-dependent magnetization of the system was then estimated by performing standard Metropolis Monte-Carlo (MC) simulations on top of the DFT-calculated magnetic exchange parameters, $J_n$'s. Regarding the MC simulations, 
$4\times4\times1$ lattice with periodic boundary conditions in the lateral 
directions (out-of-plane periodicity was considered for bulk) was constructed. 
Randomly generated spin configuration was cooled down from $2500$ to $0$ K, 
where at each temperature 2$\times$10$^3$ spin-flips per site were executed 
to obtain thermal equilibrium.

\begin{table}[b]
\caption{\label{main} In-plane lattice constants, $a$ and $b$; out-of-plane 
lattice constant for bulk, $c$; the average distance between Fe$^{\text{out}}$ 
atoms per number of layers, $d$; the electron donation per Fe atom, 
$\varDelta\rho$, the first and second value discriminating Fe$^{\text{in}}$ and 
Fe$^{\text{out}}$, respectively; the magnetic ground state, MS; the 
cohesive energy per unit formula, $E_{\text{c}}$; the energy band gap (GGA with 
$U = 4$ eV), $E_{\text{gap}}$.}
\begin{tabular}{lcccccccccccccccc}
\hline\hline
     & $a$     & $b$   & $c$   & $d$   &$\varDelta\rho$  &MS &$E_{\text{c}}$ 
&$E_{\text{gap}}$  \\
     & (\AA{}) &(\AA{})&(\AA{})&(\AA{})& $(e^{-})$       &-  & (eV)          & 
(eV) \\
\hline
Bulk & $5.08$  & 5.08  & 13.72 & -  &1.8    & AFM & 5.00 & 2.10    \\
4L   & $5.10$  & 5.10  & -     & 4.12  &1.8/1.6& AFM & 4.90 & M  \\
3L   & $5.12$  & 5.12  & -     & 3.95  &1.8/1.6& AFM & 4.87 & M  \\
2L   & $5.14$  & 5.14  & -     & 3.60  &1.8/1.6& AFM & 4.81 & M  \\
1L   & $5.16$  & 5.16  & -     & 3.04  &1.8/1.6& AFM & 4.62 & 0.75   \\
\hline\hline 
\end{tabular}
\end{table}

\section{Results and Discussion}\label{results}

\subsection{Thickness-dependent structural properties}\label{struct}

The well-known bulk $\alpha$-Fe$_2$O$_3$ structure belongs to $R\overline{3}2/c$ 
space group in which ABC stacked buckled hexagonal sublayers of Fe atoms have 
octahedral coordination with six O atoms as shown in Fig. \ref{f1}(a). In-plane and out-of-plane lattice parameters are found to be $a=b=5.08$ \AA and $c=13.72$ \AA, as listed in Table 
\ref{main}. Since Balan \textit{et al.} reported that atomically thin layers exfoliated in [001] directions are more stable \cite{Balan}, we consider the thickness-dependent properties for [001] hematite samples. As an initial structure, we truncated a layer from the bulk in the 
[001] direction, as shown by dashed lines in Fig. \ref{f1}(a). The unit cell of 
the truncated layer consists of four Fe and six O 
atoms which fits the formula unit of Fe$_2$O$_3$. Differing from the bulk form, 
there are two types of Fe atoms due to the absence of the coexisting atoms in 
the out-of-plane direction: inner Fe atoms with the octahedral bond 
coordination, labeled Fe$^{\text{in}}$, and the outer Fe atoms with the trigonal 
pyramidal coordination, labeled Fe$^{\text{out}}$ (see Fig. \ref{f1}(a)).

The optimized 1L structure (hematene), shown in Fig. \ref{f1}(a), has a 
perfectly symmetric hexagonal sublattice of Fe$^{\text{in}}$ atoms sandwiched by 
directly stacked two trigonal sublattices of Fe$^{\text{out}}$ atoms. The 
octahedral coordination of Fe$^{\text{in}}$ and trigonal pyramidal coordination 
of Fe$^{\text{out}}$ atoms remain such. The space group of the new 2D structure 
is $P\overline{3}12/m$. As given in Table \ref{main}, the lattice parameters are 
found to be $a=b=5.16$ \AA, slightly larger than those of bulk. The distance 
between the Fe$^{\text{out}}$ atoms is $d=3.04$ \AA, smaller than $3.95$ \AA in 
bulk. It is also found that each Fe$^{\text{in}}$ atom donates 
1.8 $e^{-}$ to an O atom. Fe$^{\text{out}}$ donates 1.6 $e^{-}$ which is 
slightly smaller due to having coordination 3 instead of 6. 

Differing from the monolayer, the structural configurations and the bond 
coordination of the thicker systems, 2L, 3L, and 4L, remain similar to those of 
bulk $\alpha$-Fe$_2$O$_3$, as shown in Fig. \ref{f1}(b). Slight structural 
changes presented in Table \ref{main} indicate that adjacent layer(s) in the 
out-of-plane direction prevent atoms from rearranging their position as obtained 
in the optimization of the 1L structure.

\begin{figure*}[t]
\includegraphics[width=0.9\linewidth]{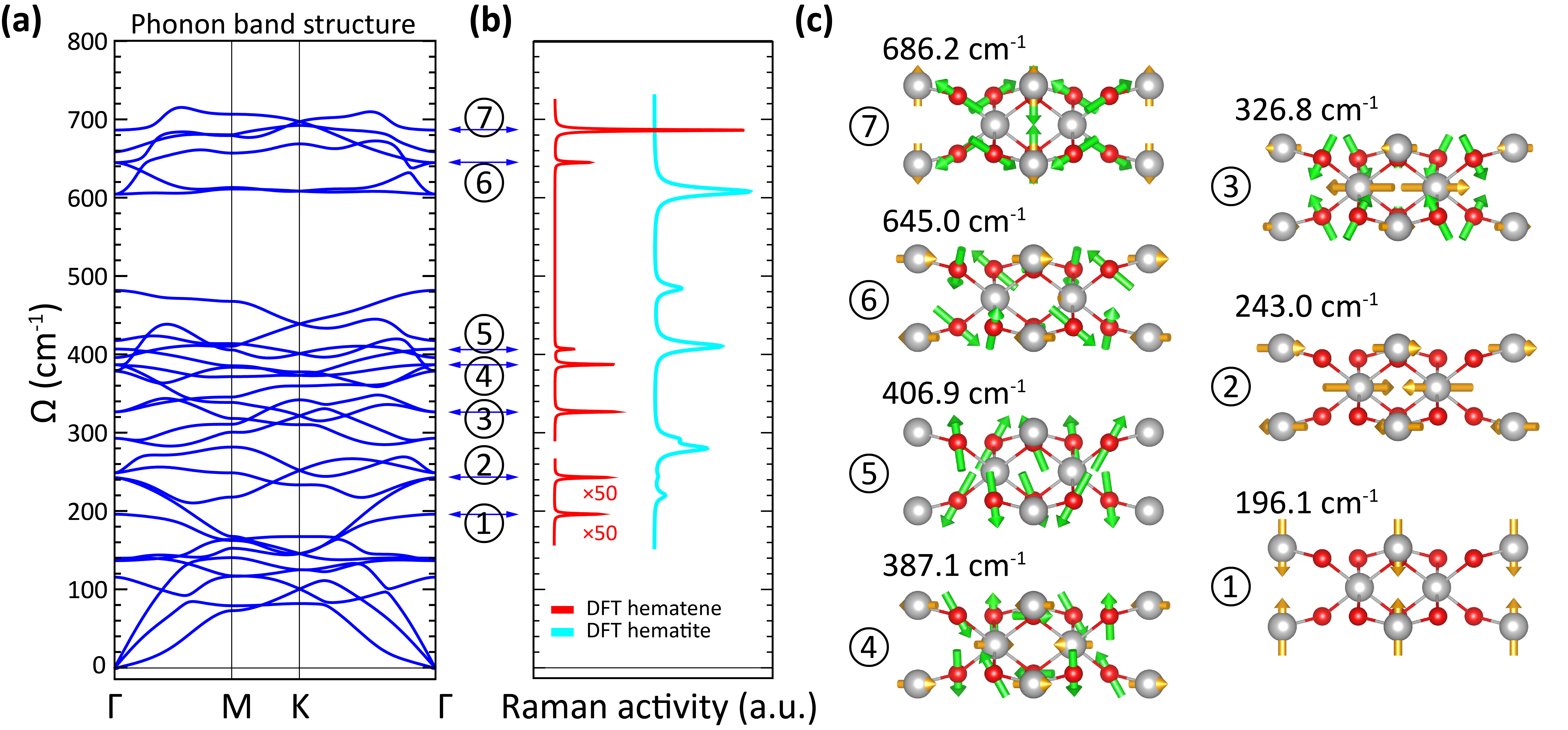}
\caption{\label{f2}
(color online) (a) Phonon band structure of hematene through high symmetry 
points of the BZ. (b) Calculated Raman spectrum for hematene and bulk 
hematite (red and turquoise, respectively). (c) The vibrational characteristics 
of Raman active phonon modes of hematene.}
\end{figure*}

\subsection{Thickness-dependent vibrational properties}\label{vibr}

The dynamical stability of the optimized 1L $\alpha$-Fe$_2$O$_3$ is examined by 
calculating its phonon band dispersion through the high symmetry points of the 
Brillouin zone (BZ), as shown in Fig. \ref{f2}(a). It is evident that hematene 
is dynamically stable, and exhibits thirty phonon branches - three of which are 
acoustic. The highest frequency of the optical phonon branches at the 
$\Gamma$ point is found to be 686.2 cm$^{-1}$, which is relatively large as 
compared to those of MX$_2$ where M=Mo, W, Re and X=S, Se (401.0$-$450.0 
cm$^{-1}$)\cite{raman-myk}. Having optical phonon branches at higher frequency 
indicates the relatively high mechanical stiffness of 1L $\alpha$-Fe$_2$O$_3$. 
For illustrative comparison, we also give the values of graphene (1555.0 
cm$^{-1}$) and $h$-BN (1343.4 cm$^{-1}$).

For further elucidation of the vibrational properties, the first order 
off-resonant Raman activities of the phonon modes for 1L $\alpha$-Fe$_2$O$_3$ 
are calculated and presented in Fig. \ref{f2}(b). There are seven Raman active 
phonon modes which are assigned as 1-to-7 from low to high frequencies (see Fig. 
\ref{f2}(c)). The seven Raman active modes can be categorized as three 
non-degenerate out-of-plane (having frequencies 196.1, 406.9, and 686.2 
cm$^{-1}$) and four doubly-degenerate in-plane vibrational modes (with 
frequencies of 243.0, 326.8, 387.1, and 645.0 cm$^{-1}$). The mode at frequency 
196.1 cm$^{-1}$ has the lowest Raman activity and is attributed to 
the purely out-of-plane breathing-like vibration of Fe atoms (see Fig. 
\ref{f2}(c)). Another non-degenerate phonon mode at the frequency of 406.9 
cm$^{-1}$ demonstrates the mostly out-of-plane vibration of O atoms while Fe 
atoms make no contribution. The highest frequency optical mode is found at 686.2 
cm$^{-1}$, which is the most prominent peak and represents the purely 
out-of-plane vibration of Fe atoms and mostly in-plane vibration of O atoms. 
Notably, this phonon mode can only be observed in 2D $\alpha$-Fe$_2$O$_3$ 
structures and is absent in bulk (or unobservable due to its negligibly small 
Raman activity). The doubly-degenerate phonon mode at 243.0 cm$^{-1}$ has a low 
Raman activity and it is attributed to the in-plane vibration of Fe atoms 
against each other, while O atoms do not contribute to the vibration. The mode 
at 326.8 cm$^{-1}$ arises from the in-plane opposite vibration of Fe atoms while 
O atoms vibrate in opposite out-of-plane directions at the top and the bottom of 
the layer. A similar vibrational character is found for the phonon mode at 
frequency 387.1 cm$^{-1}$, in which adjacent O atoms vibrate in opposite 
out-of-plane directions. Finally, the phonon mode at 645.0 cm$^{-1}$ has similar 
vibrational character with the mode at frequency 387.1 cm$^{-1}$. In this mode, 
the Fe$^{\text{in}}$ atoms have relatively small contribution to the vibration.

The calculated Raman spectrum of bulk $\alpha$-Fe$_2$O$_3$ reveals also seven 
Raman active peaks. Two out-of-plane non-degenerate phonon modes are 
assigned as A$_{1g}$ modes at the frequencies of 220.0 and 484.3 cm$^{-1}$. Other five Raman active modes having frequencies 244.0, 280.0, 292.1, 410.4, and 
607.9 cm$^{-1}$ are doubly-degenerate and are assigned as E$_g$ modes.

\begin{figure}[b]
\includegraphics[width=\linewidth]{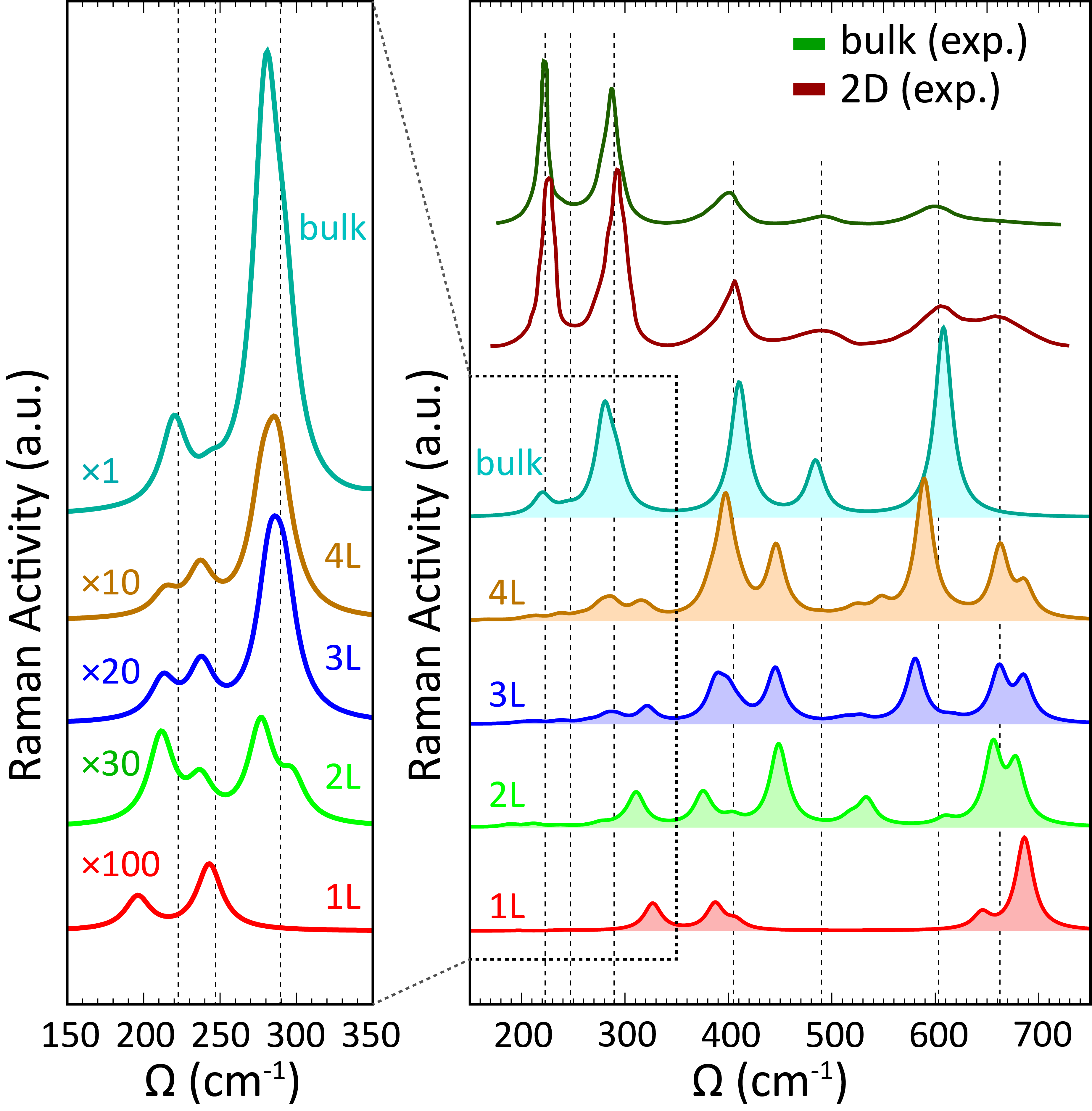}
\caption{\label{f3} (color line) Calculated Raman spectrum for 1L, 2L, 3L, 4L, and bulk. Above them, the experimental Raman spectrum for bulk and 2D 
$\alpha$-Fe$_2$O$_3$ is shown, as reported in Ref. \onlinecite{Balan}. The left panel zooms in on the low-intensity peaks in the frequency range 150-350 cm$^{-1}$. The intensities are multiplied by a given factor to enhance the visibility of the peaks.}
\end{figure}

In Fig. \ref{f3}, we present our results on the Raman spectra of 
$\alpha$-Fe$_2$O$_3$ structures from monolayer to bulk, with comparison to the experimental observations reported by 
Balan \textit{et al.}\cite{Balan}. Since the intensities of the peaks are relatively low at the frequencies below 350 cm$^{-1}$, we zoom in on the spectra in 150-350 cm$^{-1}$ range in the left panel of the figure, where peak-intensities are multiplied by given factors to facilitate the observation of the frequency-shift. The dashed lines track the peak-positions with respect to the experimental peak-positions on the frequency axis. In general, the calculated Raman spectrum of the bulk shows good agreement with the experiment of Ref. \onlinecite{Balan}. For the 2-4L samples, the calculated Raman spectra capture the main features of the experimental spectrum of a quasi-2D structure, with slightly stronger shift of the peak frequencies as compared to the experimental shift upon exfoliation. However, monolayer Fe$_2$O$_3$ exhibits distinctive peak positions. Balan \textit{et al.} demonstrated in their thickness measurement the absence of samples as thin as a monolayer of either [001] (3.04 \AA) or [010] (3.09 \AA) phases; our results for Raman response of a monolayer corroborate that fact, as experimentally measured Raman spectrum is completely different from the Raman spectrum expected for a monolayer.

We start the discussion with the peaks in the range 150-350 cm$^{-1}$, zoomed out in the left panel of Fig. \ref{f3}. The frequency of the A$_g$ 
mode displays phonon hardening with increasing thickness. Its 
frequency in a monolayer (196.1 cm$^{-1}$) hardens to 211.5 cm$^{-1}$ in 2L 
$\alpha$-Fe$_2$O$_3$, and further to 212.7 and 214.3 cm$^{-1}$ in 3L and 4L 
structures, respectively. Since this mode is attributed to the vibration of 
Fe$^{\text{out}}$ atoms, the change in its frequency decreases with increasing thickness. Therefore, for sufficiently thick samples its frequency is very close 
to the bulk case. Notably, this behavior is rather different from a layered 
(vdW) material, because in a 
non-layered material, the coordination, length, and strength of the bonds, also 
the surface reconstructions, may change significantly with thickness. As 
compared to the experiment\cite{Balan}, our finding indicates that a very small 
frequency shift of A$_g$ mode compared to bulk is a signature of a few-layer 
thick sample. Contrarily, the bulk out-of-plane mode at frequency 484.3 
cm$^{-1}$ displays phonon softening as the structure is thinned down to 2L. Its 
frequency is found to be 457.1, 463.5, and 465.4 cm$^{-1}$ in 2L, 3L, and 4L 
structures, respectively. In the case of E$_g$ modes, phonon hardening is mostly 
found with increasing thickness. The mode at frequency 276.7 cm$^{-1}$ (in 2L 
structure) displays phonon hardening to 292.1 cm$^{-1}$ (in bulk) and it is 
attributed to the in-plane shear vibration of the Fe atoms. Moreover, the phonon 
mode having frequency 533.6 cm$^{-1}$ in 2L crystal hardens to 607.9 cm$^{-1}$ 
in bulk crystal. Notably, those two E$_g$ modes are entirely absent in the 1L 
case. On the other hand, two particularly Raman active phonon modes at high 
frequencies appear in 2D structures, while being absent in the bulk crystal. 
Those two modes are found to be attributed to the mixed in- and out-of-plane 
vibrations of O atoms while the Fe atoms vibrate in-plane or out-of-plane (modes 
6 and 7 in Fig. \ref{f2}(c)). The phonon mode at 645.0 cm$^{-1}$ in 1L hardens 
to 655.7, 661.6, and 662.6 cm$^{-1}$ in 2L, 3L, and 4L structures, respectively. 
It is clear that the increment of the phonon frequency decreases with layer 
thickness. However, the Raman activity of the phonon mode decreases as the 
structure becomes thicker, and disappears for bulk hematite. Notably, these two phonon modes exhibit distinctive features for distinguishing of 2D structures from bulk, since they predominantly arise from vibrations of surface atoms. It is rather remarkable that the phonon frequency differences between bulk and exfoliated samples reported by Balan \textit{et al.} closely resemble those in our results down to 2L structure.\footnote{It should be noted that since Balan \textit{et al.}\cite{Balan} synthesized samples by exfoliation in crystalline directions of [001] and [010], there is a possibility that their Raman spectrum has also contribution from few-layers of [010] phase, but we do not expect significant differences in Raman activity of few-layer [001] and [010] hematite.} However, we also show that [001] monolayer hematene exhibits a completely distinctive Raman spectrum from the thicker structures, which may serve as a tool to distinguish true monolayer hematene from few-layer hematite in future experiments. 

\begin{figure}[t]
\includegraphics[width=\linewidth]{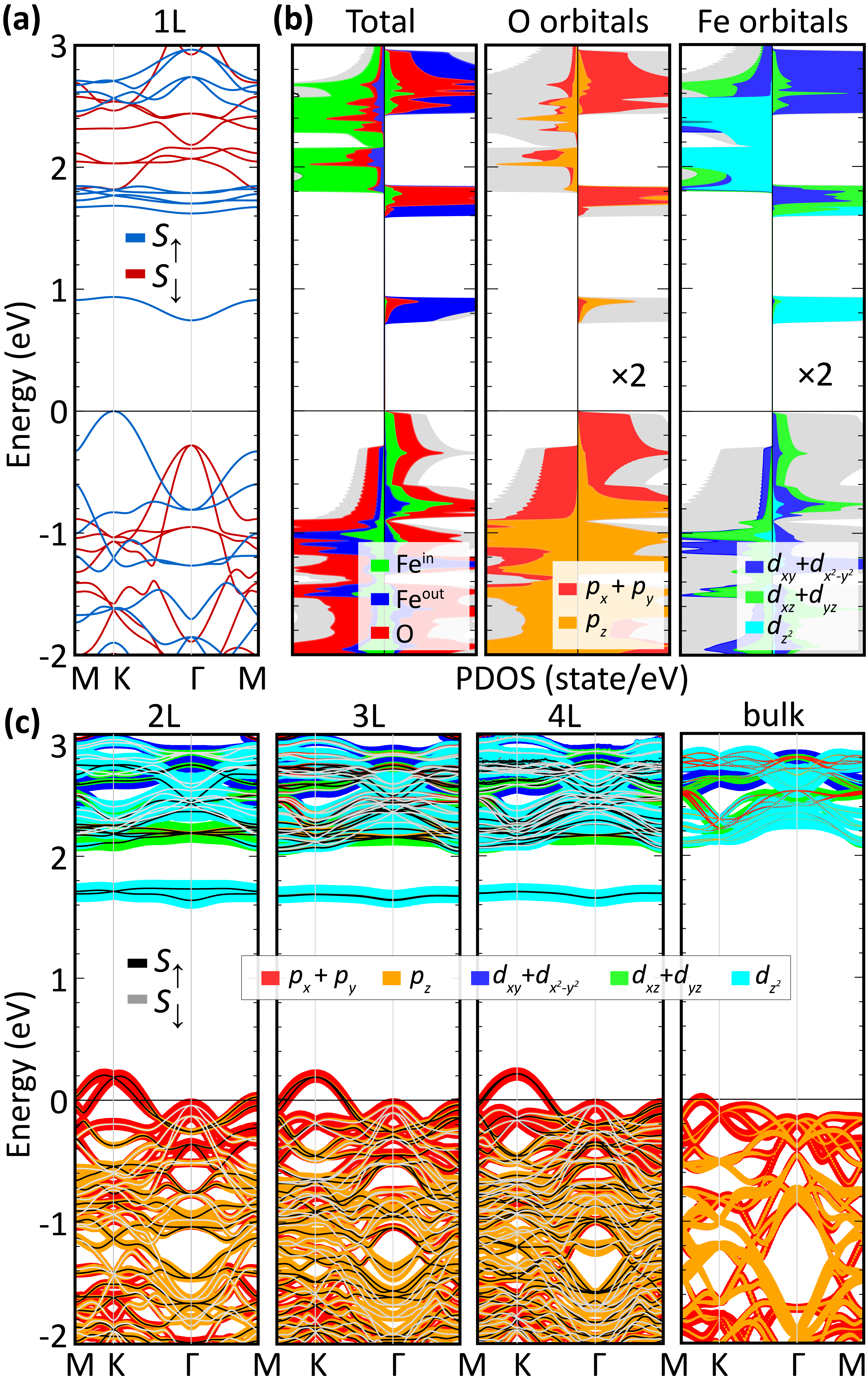}
\caption{\label{f4}(color line) (a) The spin-polarized electronic band 
structure 
of 1L $\alpha$-Fe$_{2}$O$_{3}$. (b) Atom- (left) and orbital-decomposed partial 
density of states (pDOS) of 1L $\alpha$-Fe$_{2}$O$_{3}$ , with color codes 
given 
in the graphs. The grey regions 
are the total density of state given as a reference for 
comparison. (c) Orbital-decomposed electronic band structure of 2L, 3L, 4L, and 
bulk $\alpha$-Fe$_{2}$O$_{3}$.}
\end{figure}

\subsection{Thickness-dependent electronic properties}\label{electr}

The calculated electronic band structure of 1L $\alpha$-Fe$_{2}$O$_{3}$ is shown 
in Fig. \ref{f4}(a). The band structure has asymmetric dispersion depending on the spin component. The blue curve (denoting $up$-spin, 
\textit{S}$_{\uparrow}$), has its valence band maximum (VBM), which 
is also global maximum, at the $K$ point. The 
\textit{S}$_{\uparrow}$ component shows mid-gap states, dispersive over the 
energy axis between $0.75-0.94$ eV in all directions of the BZ, indicating that 
the mid-gap band is conductive with minimum at the $\Gamma$ point. Therefore, 
the conduction band minimum (CBM) of the \textit{S}$_{\uparrow}$ component is 
the global minimum and makes hematene an indirect band-gap semiconductor with a 
band gap of 0.75 eV. In addition, regardless of the mid-gap states, the 
conduction 
band of the \textit{S}$_{\uparrow}$ component has a minimum also at the $K$ 
point with energy of 1.62 eV. Note that all extreme points including the 
mid-gap 
band originate from a single spin component of \textit{S}$_{\uparrow}$.

On the other hand, the $down$-spin (\textit{S}$_{\downarrow}$) component, with 
corresponding red lines in Fig. \ref{f4}(a), possesses its valence band maximum 
with two degenerate light and heavy hole bands at the $\Gamma$ point. The VBM of 
the \textit{S}$_{\downarrow}$ component appears at the $M$ point with an energy 
of $-0.28$ eV with respect to Fermi-level. The conduction band of the 
\textit{S}$_{\downarrow}$ component at the $M$ point is at 1.82 eV, giving rise 
to a band gap of \textit{S}$_{\downarrow}$ component of 2.10 eV.

For further analysis of the electronic properties, the atom- and 
orbital-decomposed partial density of states (pDOS) of 1L 
$\alpha$-Fe$_{2}$O$_{3}$ was calculated. In the left panel of Fig. \ref{f4}(b), 
the atom-decomposed pDOS for Fe$^{\text{in}}$, Fe$^{\text{out}}$, and O atoms is 
shown. The VBM, in which only the \textit{S}$_{\uparrow}$ states exist, is 
dominated by the oxygen states and considerable amount of states originate from 
only Fe$^{\text{in}}$ atoms. For the \textit{S}$_{\downarrow}$ part, most of the 
states are from oxygen, yet only the states of Fe$^{\text{out}}$ atoms have 
contributions. The mid-gap band, on the other hand, is dominated by the 
Fe$^{\text{out}}$ states and there is a small contribution from the oxygen 
states. 

In central and right panels of Fig. \ref{f4}(b), the pDOS for $p$-orbitals of O atoms and $d$-orbitals of Fe atoms are shown, respectively. It is revealed that the O domination of the VBM (central panel), originates from the $p_{x}$ and $p_{y}$ orbitals, while most of the Fe states (right panel) stem from in- and out-of-plane hybrid orbitals of $d_{xz}$ and $d_{yz}$, with small contributions from $d_{xy}$ and $d_{x^{2}-y^{2}}$ orbitals. The \textit{S}$_{\downarrow}$ part of the valence band has a similar picture considering the O orbitals, however, the only contribution of Fe atoms are from $d_{xy}$ and $d_{x^{2}-y^{2}}$ orbitals.

The mid-gap states, on the other hand, originate mostly from the $d_{z^{2}}$ 
orbitals of Fe$^{\text{out}}$ atoms, as shown in left and right panels of Fig. \ref{f4}(b), which is a typical example of the conductive surface states with a single chanel. The higher energies of the CB consist of the combination of 
$p$-orbitals of O atoms and $d$-orbitals o Fe.

In order to describe the change of the electronic properties with thickness, we calculate the orbital-decomposed electronic band structures of 2-4L and bulk 
$\alpha$-Fe$_{2}$O$_{3}$ shown in Fig. \ref{f4}(c). It is clearly seen that all 
few-layer structures have bands crossing the Fermi level which makes them 
metallic. These bands are dominated by $p_{x}$ and 
$p_{y}$ orbitals of O atoms. Note that in order to obtain accurate occupancies of the states around the Fermi-level in a layer exfoliated from a non-van der Waals material, the smearing parameter should be carefully chosen. If the smearing parameter is inappropriately large, the dangling bonds can be occupied and the Fermi-level can be shifted into the band gap, leading to incorrect conclusions about the electronic properties of the sample.\footnote{It should also be noted that in calculation of electronic properties of a real semiconductor, independently from the smearing parameter, the bands are fully occupied and Fermi level is always in the band gap.} Therefore, we decreased our smearing parameter down to 0.01 eV, which allowed us to capture the dangling bonds of O atoms in 2-4L samples. 

Regarding the higher energy features of the band structures, all few-layer systems exhibit mid-gap bands, dominated mainly by $d_{z^{2}}$ orbitals of Fe atoms, and the spin states are split. The bulk system, on the other hand, has degenerate spin states in the band structure. It is clear that the anisotropic spin channels, the mid-gap states and the states crossing the Fermi level in few-layer structures are a consequence of the surface.

\subsection{Thickness-dependent optical properties}\label{optical}

\begin{figure}[t]
\includegraphics[width=0.8\linewidth]{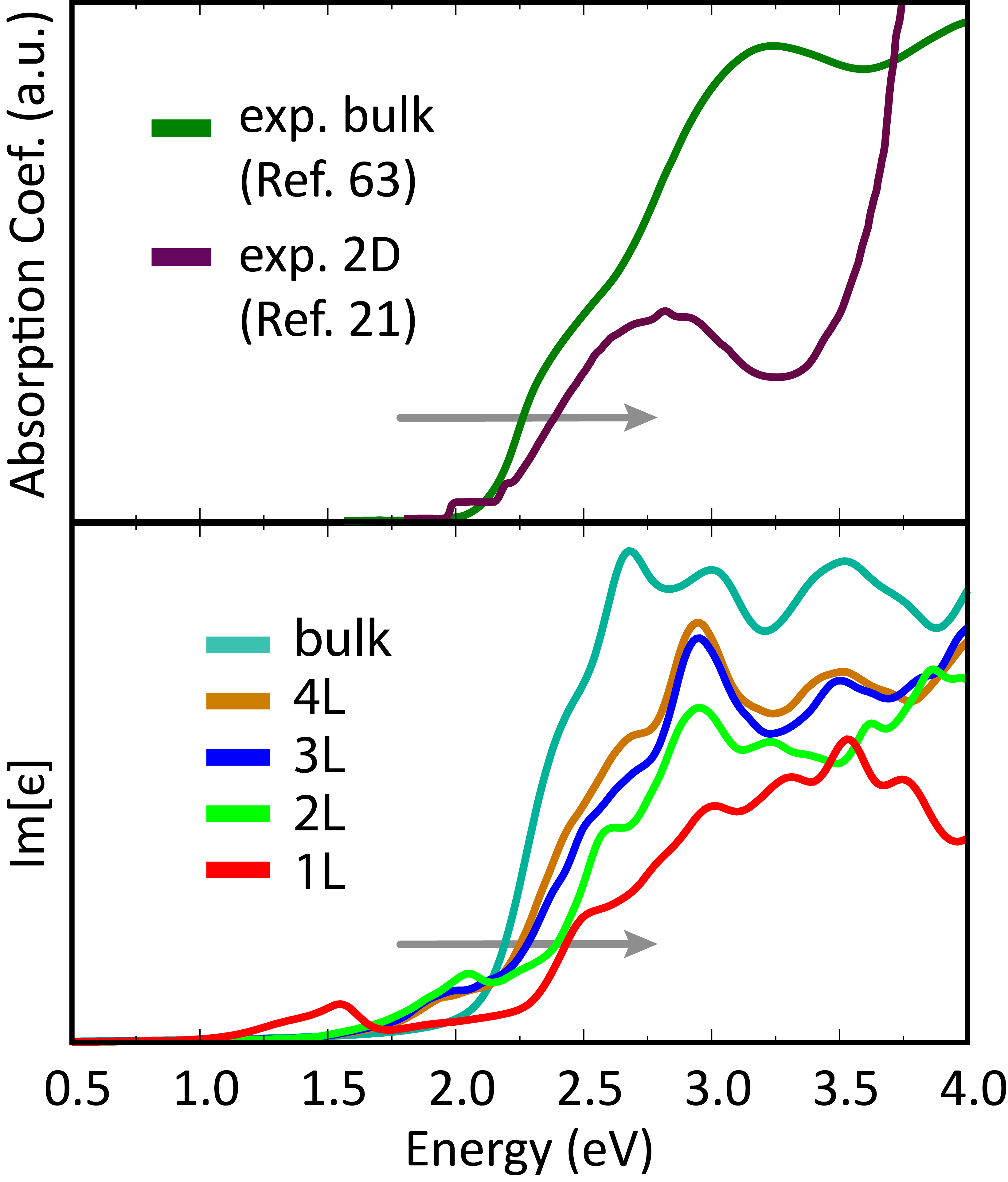}
\caption{\label{f5} (color online) Top panel: The experimentally-obtained 
absorption coefficients for bulk and 2D sample. Bottom panel: Calculated 
dielectric functions of 1L, 2L, 3L, 4L, and bulk. Arrow indicates the direction 
of change with thinning the sample.}
\end{figure}

In order to investigate the optical properties, we calculated the dielectric 
function of bulk and 1-4L $\alpha$-Fe$_{2}$O$_{3}$, by considering the direct 
excitation between the single-particle states. Then we compare the results with 
the experimental measurements on bulk\cite{Hayes} and 2D\cite{Balan} samples, 
shown in Fig. \ref{f5}. Note that the optical gaps taken from the measured and 
from the calculated optical spectra for a bulk system \textit{perfectly match}, 
which yields confidence in the theoretical methodology to directly and reliably 
compare the calculated spectra of 1-4L systems with the experimentally obtained 
spectrum of a 2D sample.\cite{Balan} 

In general, the experimental spectra of bulk and 2D structure exhibit two main 
features. (i) When the structure thinned down from bulk to 2D, the prominent 
features shift to higher energy. That shift is also clearly seen in our 
calculated spectra. (ii) 2D spectrum exhibits a small step-like feature at the 
lower-energy edge of the prominent peak, while the bulk peak rises abruptly - 
which are also confirmed by the calculated spectra.

Looking deeper in the data, the calculated spectrum of hematene (red line) 
displays its lowest transition at $\sim$1.5 eV, which is only possible when the 
mid-gap states of 1L shown in Fig. \ref{f4} are optically active. From pDOS 
analysis, we infer that the peak at 
$\sim$1.5 eV stems mainly from the optical transitions between $p_{x}$-$p_{y}$ 
states of oxygen atoms at $\sim$-0.4 eV, and $d_{z^{2}}$ of Fe$^{\text{out}}$ at 
$\sim$0.75$-$0.94, which correspond to mid-gap band. Recalling the results of 
the previous section, it is revealed here that those transitions are possible 
for the single-spin state. With its distinct mid-gap peak, the calculated 
spectrum for hematene does not match the experimentally seen absorption spectrum 
for 2D $\alpha$-Fe$_{2}$O$_{3}$ in Ref.~\onlinecite{Balan}, corroborating the 
fact that the 2D structure in experiment was thicker. As shown in our data, the 
thicker structures of 2-4L exhibit step-like feature at $\sim$1.9 eV, coinciding 
with the lower energy edge of the prominent peak. Similarly to mid-gap peak of 
1L, those step-like features originate from the transitions between occupied 
$p_{x}$ and $p_{y}$ orbitals of oxygen and mid-gap $d_{z^{2}}$ orbital of 
surface Fe atoms. Note that for 2-4L structures there are bands which are 
crossing the Fermi-level due to dangling bonds of O atoms which is 
shown in Fig. \ref{f4}(c), 
however, there is no optical transition found from the occupied bands to those 
dangling states which are just above the Fermi-level. Therefore, 2-4L 
$\alpha$-Fe$_{2}$O$_{3}$ has optical band gap, even though it has metallic 
features originating from optically inactive surface states. 

This optical spectrum analysis therefore reveals the clear distinction between 
the spectra of few-layer and monolayer samples, and also between few-layer and 
bulk hematite. Our calculations also confirm that experimental realization of 2D 
$\alpha$-Fe$_{2}$O$_{3}$ to date was in 2L and thicker forms, and not a 
monolayer.

\subsection{Thickness-dependent magnetic properties}\label{magnet}

\begin{figure*}[ht]
\includegraphics[width=\linewidth]{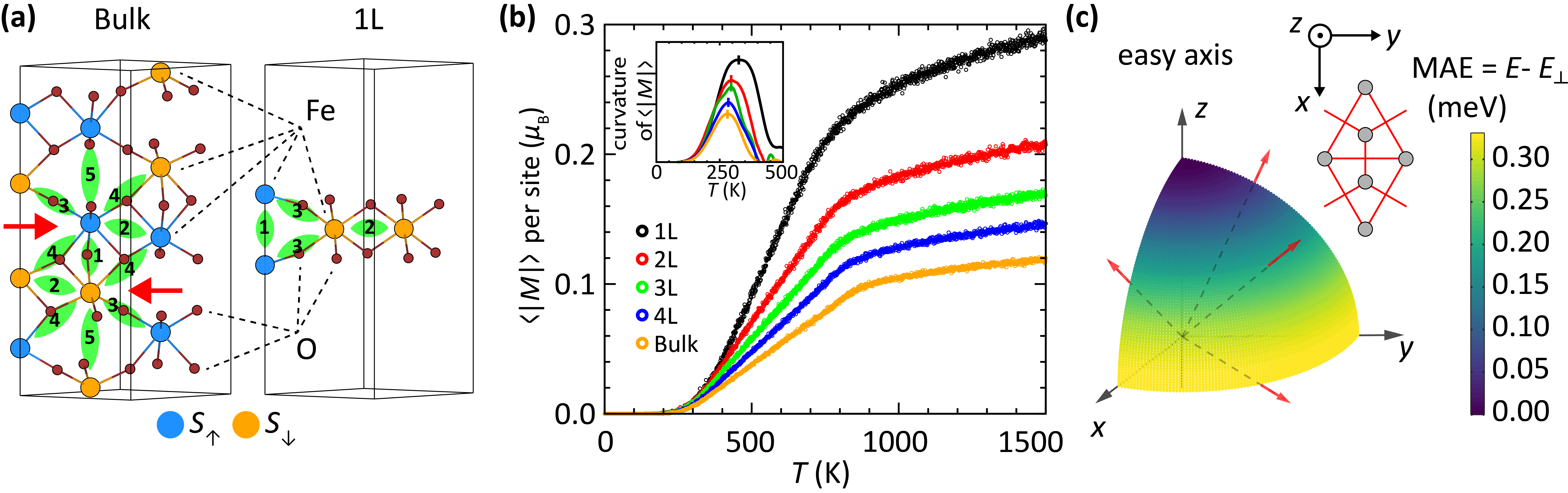}
\caption{\label{f6}(color line) (a) Schematic illustration of the magnetic 
exchange parameters between Fe atoms for bulk (left) and 1L (right). Shown
numbers are the indices of the exchange coefficients $J_n$ in Eq. \eqref{ham}. 
(b) The averaged absolute magnetization of 1-4L and bulk systems as a function 
of temperature. Inset shows $T_c$ as a function of thickness, obtained as the temperature of the maximal curvature of the corresponding magnetization curve. (c) The magnetic anisotropy energies of monolayer hematene are projected on a spherical surface, at which normal vectors correspond to magnetization direction, as shown by the arbitrarily placed red arrows.}
\end{figure*}

In this section, we discuss the thickness-dependent magnetic properties of 1-4L 
and bulk $\alpha$-Fe$_{2}$O$_{3}$. We first investigate the ground-state 
magnetic states for all structures by comparing the energetics of the different 
possible magnetic configurations. Each Fe atom has the magnetic moment of $|m| = 
4~\mu_{\text{B}}$ in the ground state for all structures. It is found that bulk 
$\alpha$-Fe$_{2}$O$_{3}$ has an anti-ferromagnetic (AFM) ground state shown in 
Fig. \ref{f6}(a), which agrees with the earlier results in 
literature\cite{Gonzalez,Padilha}. Basically, magnetization of Fe atoms in the 
hexagonal buckled 
sublayers are parallel. The sublayers have anti-parallel magnetization with 
respect to each other. The same configuration is also found as the ground state 
for the 2-4L $\alpha$-Fe$_{2}$O$_{3}$. The state for 1L is shown in Fig. 
\ref{f6}(a) next to bulk, and is also resembling the state truncated from the 
bulk AFM state.  

\begin{table}[b]
\caption{\label{tb2} The magnetic exchange parameters for monolayer, few-layer, and bulk hematite, $J_{n}$, where $n$ 
is the index of the considered nearest neighbor. For the 2-4L systems, $J_{n}$'s vary depending on the pair position in the sample. We list them top to bottom as from outermost to innermost layer of sample. The last column gives the
obtained critical temperature, $T_{c}$.}
\begin{tabular}{lcccccccccccccccc}
\hline\hline
     &$J_{1}$ &$J_{2}$&$J_{3}$& $J_{4}$ & $J_{5}$ & $T_{c}$ \\
     & (meV)  & (meV) & (meV)&  (meV) & (meV)   & (K)   \\
\hline
Bulk &1.20    &1.31   &10.37  &7.06     &0.50     &  281    \\
\hline
4L   &0.58    &1.64   &16.72  &10.26    &-0.21    &  285    \\
     &1.05    &1.09   &10.13  &7.83     &0.64     &           \\
     &1.02    &1.45   &8.56   &6.91     &0.56     &           \\
     &        &       &       &7.01     &0.51     &           \\
     &        &       &       &6.89     &         &           \\
     &        &       &       &5.81     &         &           \\
\hline
3L   &0.47    &1.63   &16.51  &10.15    &-0.33    &  297       \\
     &0.92    &1.09   &8.29   &10.03    &0.52     &           \\
     &        &1.35   &       &6.89     &0.45     &           \\
     &        &       &       &6.63     &         &           \\
     &        &       &       &5.69     &         &           \\
\hline
2L   &0.37    &1.73   &16.44  &9.79     &-0.67    &  299     \\
     &        &0.83   &8.05   &7.32     &0.43     &           \\
     &        &       &       &5.57     &         &           \\
\hline
1L   &0.94    &2.13   &11.88  &         &         &  326    \\
\hline\hline 
\end{tabular}
\end{table}

The energetics of different magnetic configurations also allows us to calculate 
the magnetic exchange parameters between the magnetic sites. We map the 
energetics to Heisenberg spin Hamiltonian given in Eq. \eqref{ham} by applying 
the nearest-neighbor (NN) approximation following the procedure given in Sec. \ref{comp}. In the bulk structure, as depicted in Fig. \ref{f6}(a), the 
magnetic sites are identical in terms of number of $n$th nearest neighbors and the corresponding bond length. However, for 1-4L structures, the number of $n$th nearest neighbors and the bond lengths differ depending on the position of the magnetic site in question in the out-of-plane direction. Therefore, we performed 
different sets of calculations for the magnetic exchange parameters experienced by an atom, depending on its position in the sample. The obtained values are listed in Table \ref{tb2}. 

In order to obtain the magnetic properties beyond the ground state, we performed 
MC simulations based on Ising model, using the determined exchange 
parameters, to predict the temperature-dependent magnetic properties such as 
critical temperature ($T_c$). However, the 
Ising model considers only up/down collinear spins, hence yields much 
overestimated barriers for magnetic transitions. As a consequence, our trial MC 
calculation using DFT-obtained exchange parameters for the bulk system yield much larger $T_c$ for ferromagnetic to anti-ferromagnetic transition as compared to 
reality. To accommodate this issue, we rescaled all exchange parameters by a 
constant factor (0.65) so that the theoretically obtained $T_c = 281$ K is close to the experimentally measured $T_c$ of 265 K for bulk hematite\cite{Morin,Moriya}. 

For each structure, starting from $2500$ K, the system is cooled down to $0$ K, 
and the average absolute magnetic moment per magnetic site, $\langle | M | 
\rangle$, is obtained as a function of temperature, as shown in Fig. 
\ref{f6}(b). From the point where the magnetization curve exhibits maximal curvature as a function of temperature, we determine the critical temperature ($T_c$) at which the system goes from weak ferromagnetic to the anti-ferromagnetic state, as listed in Table \ref{tb2}. Of course, 
the behavior as a function of temperature is also very instructive, as will be 
discussed further on.

In general, as shown in Fig. \ref{f6}(b), the magnetization curves exhibit similar behavior for all samples, with overall magnetization decrease with sample thickness. All curves exhibit a strong curvature in the range $700-860$ K where the anti-ferromagnetic exchange interaction starts to take over. For different samples, the anti-ferromagnetic states gain dominance at temperatures $270$-$350$ K. It 
should be pointed out that our MC simulation cannot capture the N\'eel 
temperature ($T_{N}$) of 961 K\cite{Morin} for bulk hematite, due to the Ising approximation in which all spin magnetic moments are considered collinear (causing also the high-temperature ferromagnetism as discussed 
previously). 

We start the discussion from the bulk sample, where the magnetization (orange dots in Fig. \ref{f6}(b)) gradually decreases from on-site average of $\sim$0.12 
$\mu_{\text{B}}$ at 1500 K down to 0.10 $\mu_{\text{B}}$ at $854$ K, where the anti-ferromagnetic state starts to build up. $T_c$ is found to be 281 K and the magnetization at room temperature is found to be 0.004 $\mu_{\text{B}}$, which is consistent with the experimentally determined value of $0.005~\mu_{\text{B}}$.\cite{Moriya} The magnetization of 
4L structure (blue dots in Fig. \ref{f6}(b)) is slightly above the one of bulk. The antiferromagnetic interactions become apparent at $\sim$816 K and dominate below $T_c=285$ K, which is nearly the same as for bulk. The magnetization 
for 3L sample exhibits the same behavior as 4L and bulk structures, with slightly higher magnetization. The influence of the anti-ferromagnetic exchange arises at $\sim$771 K. $T_c$ is found to be 297 K, which can be considered as a 
significant jump as compared to bulk and 4L samples. The results for 2L system are also similar to thicker samples in terms of the form of the magnetization curve, with further increased magnetization compared to thicker samples. The anti-ferromagnetic exchange interaction is taking over at 790 K, and $T_c$ is found to be 299 K. 

The most significant changes are obtained when the sample is thinned 
down to a monolayer, even though the behavior of the magnetization curve still resembles the one found for thicker samples. The weak ferromagnetism is largest 
of all samples and the anti-ferromagnetic interaction rises at $\sim$711 K which is a significantly lower threshold temperature compared to 2L and 3L 
samples. The critical temperature of $T_c=326$ K is above the room temperature and the highest of all considered sample thicknesses.

Our MC simulation therefore reveals that considering $T_c$ values, bulk and 4L 
system on one side, and 2L and 3L systems on the other, are similar and can be 
considered equivalent. Monolayer hematene 
is different from thicker structures in all aspects, which is expected due to 
structural differences (cf. Fig. \ref{f1}). Here it should be noted that our results do not support the findings of 
Ref.~\onlinecite{Balan}, where FM to AFM transition was not observed for few-layer 2D 
$\alpha$-Fe$_{2}$O$_{3}$ down to low temperature. In addition to that, we speculate that in the experiment AFM phase 
could be further suppressed by the substrate effects so the system can preserve 
ferromagnetic behavior at even lower temperature. We reiterate the fact that 
true monolayer hematene exhibits much higher $T_c$ than 2-3L structures, so 
Morin transition should be experimentally observable there with lowered 
temperatures.

To complete the discussion, the magnetic anisotropy energies (MAE) of monolayer 
hematene were investigated as well. In Fig. \ref{f6}(c), the MAE is projected on 
a sphere surface with respect to magnetization direction. There, $x$, $y$, and 
$z$ directions correspond to $arm$-$chair$, $zig$-$zag$, and out-of-plane 
directions, respectively. The minimum energy direction is found to be parallel 
to the $z$, i.e. to the out-of-plane direction, and the corresponding energy is 
set to 0.0 meV. MAE is found to be 0.34 meV when the 
magnetization is parallel to the plane of the structure and does not change with 
the horizontal angle. 

\section{Conclusions}\label{Conc}
In summary, motivated by the recent experimental synthesis of few-layer thick 
hematite, i.e. 2D $\alpha$-Fe$_2$O$_3$\cite{Balan}, we investigated 
theoretically the structural, vibrational, magnetic, and electronic properties 
of monolayer, few-layer and bulk $\alpha$-Fe$_2$O$_3$. The monolayer 
$\alpha$-Fe$_2$O$_3$, for which we reserved the name hematene, was shown to 
indeed be stable, although not yet seen experimentally. Our calculated Raman 
spectrum shows that hematene can be clearly distinguished from few-layer and 
thicker structures, since the frequency shifts (weak when thinning the sample) 
are found to be the largest between monolayer and two-layer samples. We also 
find that all 2D structures exhibit distinctive 
Raman active modes in high frequencies that are absent in bulk. 

In the electronic band structure, hematene exhibits strong spin asymmetry. 
Valence-band maximum (VBM), mid-gap bands and conduction-band minimum all belong 
to one spin state. VBM is dominated by $p_{x}$ and $p_{y}$ orbitals of oxygen, 
while the mid-gap band originates mostly from the $d_{z^{2}}$ orbital of surface 
iron atoms. 

Our calculated optical spectra of monolayer, few-layer, and bulk structures show 
that the prominent part of the optical spectrum shifts to higher energy when the 
system is thinned down, as seen in experiment. 2-4 layer structures also exhibit 
a small step at the onset energy of the prominent peak, also seen in experiment, 
originating from the transition between the $p_{x}$ and $p_{y}$ of oxygen atoms 
to $d_{z^{2}}$ of surface iron. The similar type of transition is responsible 
for the mid-gap peak of a monolayer structure, whose separation from the 
prominent peak is another feature to distinguish true hematene from few-layer 
samples. 
 
Finally, our magnetic simulations based on DFT-determined magnetic exchange 
parameters and subsequent Monte-Carlo simulations show that all considered 
structures display transitions from weak 
ferromagnetic to anti-ferromagnetic state with decreasing temperature. 
We note however that monolayer 
hematene exhibits higher $T_c$ in our calculations, hence there the transition 
to AFM state should be easier to detect experimentally.

Owing to its tunability with atomistic changes in thickness, magnetism in a 
broad temperature range (including room temperature), optical bandgap in the 
visible range, as well as the spin-dependent electronic and optical properties, 
2D $\alpha$-Fe$_2$O$_3$ puts itself forward as a candidate for diverse 
technological applications. Taking all our findings at face value, we conclude 
that 2D $\alpha$-Fe$_2$O$_3$ is certainly worth of further investigation and use 
in functional heterostructures. 

\begin{acknowledgments}
This work was supported by Research Foundation-Flanders (FWO-Vlaanderen). 
Computational resources were provided by Flemish Supercomputer Center (VSC), and TUBITAK ULAKBIM, High Performance and Grid Computing Center (TR-Grid 
e-Infrastructure). Part of this work was also supported by FLAG-ERA project 
TRANS-2D-TMD and TOPBOF-UAntwerp. M. Y. was supported by a postdoctoral fellowship from the Flemish Science Foundation (FWO-Vl). 
\end{acknowledgments}


\begin{thebibliography}{99}

\bibitem{Novo1} K. S. Novoselov, A. K. Geim, S. V. Morozov, D. Jiang, Y. Zhang, 
S. V. Dubonos, I. V. Grigorieva, and A. A. Firsov, Science \textbf{306}, 666 
(2004).

\bibitem{Geim1}K. S. Novoselov, A. K. Geim, S. V. Morozov, D. Jiang, M. I. 
Katsnelson, I. V. Grigorieva, S. V. Dubonos, and A. A. Firsov,
Nature \textbf{438}, 197 (2005).

\bibitem{Sahin3} H. Sahin, S. Cahangirov, M. Topsakal, E. Bekaroglu, E. Akturk, 
R. T. Senger, and S. Ciraci, Phys. Rev. B \textbf{80}, 155453 (2009).

\bibitem{Wang2} Q. Wang, Q. Sun, P. Jena, and Y. Kawazoe, ACS Nano \textbf{3}, 
621 (2009).

\bibitem{Kim} K. K. Kim, A. Hsu, X. Jia, S. M. Kim, Y. Shi, M. Hofmann, D. 
Nezich, J. F. Rodriguez-Nieva, M. Dresselhaus, T. Palacios, and J. Kong, Nano 
Lett. \textbf{12} (1), 161 (2012).

\bibitem{Tsipas} P. Tsipas, S. Kassavetis, D. Tsoutsou, E. Xenogiannopoulou, 
E. Golias, S. A. Giamini, C. Grazianetti, D. Chiappe, A. Molle, M. Fanciulli, 
and A. Dimoulas, Appl. Phys. Lett. \textbf{103}, 251605 (2013).

\bibitem{Bacaksiz} C. Bacaksiz, H. Sahin, H. D. Ozaydin, S. Horzum, R. T. 
Senger, and F. M. 
Peeters, Phys. Rev. B \textbf{91}, 085430 (2015). 

\bibitem{Gordon} R. A. Gordon, D. Yang, E. D. Crozier, D. T. Jiang, and R. F. 
Frindt, Phys. Rev. B 
\textbf{65}, 125407 (2002).

\bibitem{Coleman} J. N. Coleman, M. Lotya, A. O'Neill, S. D. Bergin, P. J. 
King, 
U. Khan, K. Young, 
A. Gaucher, S. De, R. J. Smith, I. V. Shvets, S. K. Arora, G. Stanton, H. Y. 
Kim, K. Lee, G. T. Kim, G. S. Duesberg,
 T. Hallam, J. J. Boland, J. J. Wang, J. F. Donegan, 
J. C. Grunlan, G. Moriarty, A. Shmeliov, R. J. Nicholls, J. M. Perkins, E. M. 
Grieveson, K. 
Theuwissen, D. W. McComb, P. D. Nellist, and V. Nicolosi, Science \textbf{331}, 
568 (2011). 

\bibitem{Wang1} Q. H. Wang, K. Kalantar-Zadeh, A. Kis, J. N. Coleman, and M. S. 
Strano, Nat. 
Nanotech. \textbf{7}, 699 (2012).

\bibitem{Ross} J. S. Ross, P. Klement, A. M. Jones, N. J. Ghimire, J. Yan, D. 
G. 
Mandrus, T. 
Taniguchi, K. Watanabe, K. Kitamura, W. Yao, D. H. Cobden, and X. Xu, Nat. 
Nanotech. \textbf{9}, 
268 (2014).

\bibitem{Sahin2} H. Sahin, S. Tongay, S. Horzum, W. Fan, J. Zhou, J. Li, J. Wu, 
and F. M. Peeters, 
Phys. Rev. B \textbf{87}, 165409 (2013).

\bibitem{Tongay} S. Tongay, H. Sahin, C. Ko, A. Luce, W. Fan, K. Liu, J. Zhou, 
Y. S. Huang, C. H. Ho, J. Yan, D. F. Ogletree, S. Aloni, J. Ji, S. Li, J. Li, 
F. 
M. Peeters, and 
J. 
Wu, Nat. Comm. 
\textbf{5}, 3252 (2014). 

\bibitem{Horzum} S. Horzum, D. Cakir, J. Suh, S. Tongay, Y. S. Huang, C. H. Ho, 
J. 
Wu, H. Sahin, and F. M. Peeters, 
Phys. Rev. B \textbf{89}, 155433 (2014).

\bibitem{Chen3} B. Chen, H. Sahin, A. Suslu, L. Ding, M. I. Bertoni, F. M. 
Peeters, and S. Tongay, 
ACS Nano \textbf{9} (5), 5326 (2015).

\bibitem{Cahangirov} S. Cahangirov, M. Topsakal, E. Akturk, H. Sahin, and S. 
Ciraci, Phys. Rev. 
Lett. \textbf{102}, 236804 (2009). 

\bibitem{Kara} A. Kara, H. Enriquez, A. P. Seitsonen, L. C. L. Y. Voon, S. 
Vizzini, B. Aufray, and 
H. Oughaddou, Surf. Science Report. \textbf{67}, 1 (2012).

\bibitem{Ga_N6} V. Kochat, A Samanta, Y. Zhang, S. Bhowmick, P. Manimunda, S. 
Asif, A. S. Stender, R. Vajtai, A. K. Singh, C. S. Tiwary, and P. M. Ajayan,
Sci. Adv. \textbf{4}, e1701373 (2018).

\bibitem{ga1} S. V. Badalov, M. Yagmurcukardes, F. M. Peeters, and H. Sahin, J. 
Phys. Chem. C \textbf{122}, 28302 (2018).

\bibitem{ga2} M. Nakhaee, M. Yagmurcukardes, S. A. Ketabi, and F. M. Peeters, 
Phys. Chem. Chem. Phys. \textbf{21}, 15798 (2019).

\bibitem{Balan} A. P. Balan, S. Radhakrishnan, C. F. Woellner, S. K. Sinha,
L. Deng, C. de los Reyes, B. M. Rao, M. Paulose, R. Neupane,
A. Apte, V. Kochat, R. Vajtai, A. R. Harutyunyan, C.-Wu Chu, G. Costin,
D. S. Galvao, A. A. Marti, P. A. van Aken, O. K. Varghese, C. S. Tiwary,
A. M. M. R. Iyer, and P. M. Ajayan, 
Nat. Nanotech. \textbf{13}, 602 (2018).

\bibitem{Kan} E. Kan, M. Li, S. Hu, C. Xiao, H. Xiang, and K. Deng, J. Phys. 
Chem. Lett. \textbf{4}, 1120 (2013).

\bibitem{Marelli} M. Marelli, A. Naldoni, A. Minguzzi, M. Allieta, 
T. Virgili, G. Scavia, S. Recchia, R. Psaro, and V. 
Dal Santo, ACS Appl. Mater. Interfaces \textbf{6}, 11997 (2014).

\bibitem{Mishra} M. Mishra and D.-M. Chun, Appl. Catal. A \textbf{6}, 126 
(2015).

\bibitem{Chen} J. Chen, L. Xu, W. Li, and X. Gou, Adv. Mater. \textbf{17}, 582 
(2005).

\bibitem{Zeng} H. Zeng, J. Li, J. P. Liu, and Z. L. Wang, Nature \textbf{420}, 
395 (2002).

\bibitem{Katz} J. Katz, S. C. Riha, N. C. Jeong, A. B. F. Martinson, K. Omar,
O. K. Farha, and J. T. Hupp, Coord. Chem. Rev. \textbf{256}, 2521 
(2012).

\bibitem{Morin} F. J. Morin, Phys. Rev. \textbf{78}, 819 (1950).

\bibitem{Moriya} T. Moriya, Phys. Rev. \textbf{120}, 91 (1960).

\bibitem{Smith} R. D. L. Smith, M. S. Prevot, R. D. Fagan, Z. Zhang, P. A. 
Sedach, M. K. J. Siu, S. Trudel, and C. P. Berlinguette, Science \textbf{340}, 
60 (2013).


\bibitem{Sakurai} S. Sakurai, J.-I. Shimoyama, K. Hashimoto, and S. I. Ohkoshi, 
Chem.
Phys. Lett. \textbf{458}, 333 
(2008).

\bibitem{Namai} A. Namai, M. Yoshikiyo, K. Yamada, S. Sakurai, T. Goto, T.
Yoshida, T. Miyazaki, M. Nakajima, T. Suemoto, H. Tokoro, and S. I.
Ohkoshi, Nat. Commun. \textbf{3}, 
1035 (2012).

\bibitem{Ohkoshi} S. I. Ohkoshi, S. Kuroki, S. Sakurai, K. Matsumoto, K. Sato, 
and S.
Sasaki, Angew. Chem., Int. Ed. \textbf{46}, 8392 
(2007).

\bibitem{Namai2} A. Namai, S. Sakurai, M. Nakajima, T. Suemoto, K. Matsumoto,
M. Goto, S. Sasaki, and S. I. J. Ohkoshi, Am. Chem. Soc. \textbf{131}, 
1170 (2009).

\bibitem{Long} N. V. Long, Y. Yang, M. Yuasa, C. M. Thi, Y. Q. Cao, T. Nann,and 
M. Nogami, 
RSC Advances \textbf{15}, 6383 (2014). 

\bibitem{Wu} X. L. Wu, Y. G. Guo, L. J. Wan, and C. W. Hu, J. Phys. Chem. C 
\textbf{112}, 16824 (2008).

\bibitem{Banerjee} A. Banerjee, V. Aravindan, S. Bhatnagar, D. Mhamane, S. 
Madhavi, and S. Ogale, 
Nano Energy \textbf{2} (5), 890 (2013).

\bibitem{Gu} X. Gu, L. Chen, Z. C. Ju, H. Y. Xu, J. Yang, and Y. T. Qian, Adv. 
Funct. Mater. \textbf{23}, 4049 (2013).

\bibitem{Tang} W. S. Tang, Y. Su, Q. Li, S. A. Gao, and J. K. Shang, Water Res. \textbf{47}, 3624 (2013).

\bibitem{Cao} C. Y. Cao, J. Qu, W. S. Yan, J. F. Zhu, Z. Y. Wu, and W. G. Song, Langmuir \textbf{28}, 4573 (2012).

\bibitem{Wang} J. X. Wang, R. S. Yuan, L. Y. Xie, Q. F. Tian, S. Y. Zhu, Y. H. Hu, P. Liu, X. C. Shi, and D. H. Wang, RSC Advances \textbf{2}, 1112 (2012).

\bibitem{Qiu} G. H. Qiu, H. Huang, H. Genuino, N. Opembe, L. Stafford, S. Dharmarathna, and S. L. Suib, J. Phys. Chem. C \textbf{115}, 19626 (2011).

\bibitem{Dormann} J. L. Dormann, D. Fiorani, and E. Tronc, Adv. Chem. Phys. 
\textbf{98}, 
283 (1997).

\bibitem{Zysler} R. D. Zysler, D. Fiorani, A. M. Testa, M. Godinho, E. 
Agostinelli, and L. Suber, J. Magn. Magn. Mater.
\textbf{272}, 
1575 (2004).

\bibitem{Schroeer} D. Schroeer and R. C. Nininger, 
Phys. Rev. Lett.
\textbf{19}, 
632 (1967).

\bibitem{Sorescu} M. Sorescu, R. A. Brand, D. Mihaila-Tarabasanu, and L. 
Diamandescu, J. Appl. Phys.
\textbf{85}, 
5546 (1999).

\bibitem{Gonzalez} R. I. Gonzalez, J. Mella, P. D\'{\i}az, S. Allende, E. E. 
Vogel, C. Cardenas, and F. Munoz, 2D Mater. \textbf{6} 045002 (2019).

\bibitem{Padilha} A. C. M. Padilha, M. Soares, E. R. Leite, and A. Fazzio, 
J. Phys. Chem. 
C  \textbf{123}, 16359 (2019).

\bibitem{Huang} B. Huang, G. Clark, E. Navarro-Moratalla, D. R. Klein, R. 
Cheng, K. L. Seyler, D. Zhong, E. 
Schmidgall, M. A. McGuire, D. H. Cobden, W. Yao, D. Xiao, P. Jarillo-Herrero, 
and X. Xu, Nature \textbf{546}, 270 
(2017).

\bibitem{vasp1} G. Kresse and J. Hafner, Phys. Rev. B \textbf{47}, 558 (1993).

\bibitem{vasp2} G. Kresse and J. Furthmuller, Phys. Rev. B \textbf{54}, 11169 
(1996).

\bibitem{perdew} J. P. Perdew, K. Burke, and M. Ernzerhof, Phys. Rev. Lett. 
\textbf{77}, 3865 (1996).

\bibitem{grimme} S. J. Grimme, J. Comput. Chem. \textbf{27}, 1787 (2006).

\bibitem{Dudarev} S. L. Dudarev, G. A. Botton, S. Y. Savrasov, C. J. Humphreys 
and A. P. Sutton, Phys. Rev. B 
\textbf{57}, 1505 (1998). 

\bibitem{bader} G. Henkelmana, A. Arnaldssonb, H. J\'{o}nsson, Comput. Mater. 
Sci. \textbf{36}, 354 (2006).

\bibitem{phon} D. Alfe,
Comput. Phys. Commun. \textbf{180}, 2622 (2009).

\bibitem{vasp-raman} A. Fonari and S. Stauffer, 
"https://github.com/raman-sc/VASP/" (2013).

\bibitem{raman-myk} M. Yagmurcukardes, C. Bacaksiz, E. Unsal, B. Akbali, R. T. 
Senger, and H. Sahin, Phys. Rev. B 
\textbf{97}, 115427 (2018).

\bibitem{r1} M. Yagmurcukardes, F. M. Peeters, and H. Sahin,
Phys. Rev. B \textbf{98}, 085431 (2018).

\bibitem{Xiang} H. J. Xiang, E. J. Kan,  S.-H. Wei, M.-H. Whangbo, and X. G. Gong, Phys. Rev. B \textbf{84}, 224429 
(2011).

\bibitem{Hayes} D. Hayes, R. G. Hadt, J. D. Emery, A. A. Cordones, A. B. F. 
Martinson, M. L. Shelby, K. A. 
Fransted, P. D. Dahlberg, J. Hong, X. Zhang, Q. Kong, R. W. Schoenleinc, and 
L. X. Chen, Ener. Env. Sci. \textbf{9}, 
3754 (2016).




\end{thebibliography}
\end{document}